\documentclass[showpacs,twocolumn]{revtex4}
\usepackage{graphicx}
\usepackage[usenames]{color}
\usepackage{ulem}

\begin{document}

\title{From one- to two-dimensional solitons in the Ginzburg-Landau model of
lasers with frequency selective feedback}
\author{P.~V.~Paulau$^{1,2}$, D.~Gomila$^{1}$, P.~Colet$^{1}$, B.~A. Malomed$^{3}$, and W. J.~Firth$^{4}$}
\affiliation{$^{1}$ IFISC, Instituto de F\'isica Interdisciplinar y Sistemas Complejos
(CSIC-UIB), Campus Universitat Illes Balears, E-07071 Palma de Mallorca,
Spain}
\affiliation{$^{2}$ Institute of Physics, Nezavisimosty av. 68, 220072 Minsk, Belarus}
\affiliation{$^{3}$ Department of Physical Electronics, School of Electrical Engineering,
Faculty of Engineering, Tel Aviv University, 69978, Tel Aviv, Israel}
\affiliation{$^{4}$ Department of Physics, University of Strathclyde, 107 Rottenrow East,
Glasgow G4 0NG, UK}
\date{\today}
\pacs{42.65.Tg; 42.81.Dp}

\begin{abstract}
We use the cubic complex Ginzburg-Landau equation coupled to a dissipative
linear equation as a model of lasers with an external frequency-selective
feedback. It is known that the feedback can stabilize the one-dimensional
(1D) self-localized mode. We aim to extend the analysis to 2D stripe-shaped
and vortex solitons. The radius of the vortices increases linearly with
their topological charge, $m$, therefore the flat-stripe soliton may be
interpreted as the vortex with $m=\infty$, while vortex solitons can be
realized as stripes bent into rings. The results for the vortex solitons are
applicable to a broad class of physical systems. There is a qualitative
agreement between our results and those recently reported for models with
saturable nonlinearity.
\end{abstract}

\maketitle


\section{Introduction}

The field of spatial pattern formation in nonlinear systems has grown
significantly in the last decades (see reviews \cite%
{Kivshar1989,CrossHohenberg1993,Aranson2002,Etrich,Buryak2002,Desyatnikov2005,
LectNotesPhys2005,Malomed2007,Barcelona}). In particular, that
growth was significantly contributed to by the interest in
self-localized states (``solitons") and their stability in
pattern-forming systems, both conservative and dissipative ones.

The motivation of the present work is to achieve a quasi-analytical
description of the formation of stable self-localized structures in
spatially-extended lasers. To this end, we consider a complex
Ginzburg-Landau model with the cubic nonlinearity (CGL3), for which an
analytical chirped-sech localized solution is well known in the
one-dimensional (1D) setting \cite{Exact1,Exact2}. While this solution is
always unstable, it has been shown that an additional, linearly coupled,
dissipative linear equation can lead to its stabilization in
coupled-waveguide models, keeping the solution in the exact analytical form
\cite{Atai1998,Sakaguchi,Malomed2007}. Self-localized states in a wide
variety of systems described by such coupled linear and nonlinear equations,
in both 1D and 2D, was recently discussed in Ref. \cite{Firth2010}.

The physical system which offers a natural realization of such models is a
broad-area vertical-cavity surface-emitting laser (VCSEL), coupled to an
external frequency-selective feedback (FSF). This system has been a topic of
interest during the last years since it can display a variety of localized
structures on top of a non-lasing background \cite%
{Tanguy2006,TanguyPRL,Tanguy2008,Radwell2009,Radwell2010}. In this system,
the (complex) intra-VCSEL field features nonlinear spatiotemporal dynamics
due to two-way coupling between the optical field and the inversion of the
electronic population (driven by the injection current), while the feedback
field obeys a linear equation which is linearly coupled to the main equation
for the intra-VCSEL field. Previous studies have modeled the VCSEL-FSF using
various approximations and producing a rich variety of stable and unstable
localized modes, including the fundamental soliton \cite{Paulau2008} and its
complex dynamics \cite{Scroggie2009,Paulau2009}, as well as side-mode
solitons supported by an external cavity \cite{Paulau2009}, and vortex
solitons \cite{Paulau2010}.

All the above-mentioned VCSEL-FSF models have been numerically investigated
including, at various levels of the approximation, physically relevant
features, such as the feedback delay and electron-hole dynamics and
diffusion. On the other hand, the observed phenomena feature strong
similarities persisting under progressive simplifications of the model, such
as replacing the feedback grating with a Lorentzian filter, the adiabatic
elimination the electron-hole dynamics, and adopting instantaneous, rather
than delayed, feedback \cite{Paulau2010}. This observations suggest that a
simpler underlying model may be introduced. In this vein, it was implied in
Ref. \cite{Firth2010} that a simplification towards a simple cubic
approximation for the nonlinearity, could provide such a model of the CGL3
type. It was also noted that such a cubic approximation to the VCSEL-FSF
would place it in the same class of models as those previously introduced
for coupled optical waveguides with active (pumped) and passive (lossy)
cores in Refs. \cite{Winful,Atai,Atai1998,Sakaguchi,Malomed2007}, and for
pulsed fiber lasers -- in Ref. \cite{Kutz2008}.

In this work we demonstrate that such a CGL3 system does indeed allow stable
and robust fundamental solitons in 1D, and, which is the basic novel
finding, fundamental and multi-charge vortex solitons in 2D. We present the
bifurcation diagram for the fundamental 1D solitons in our generalized CGL3
model, and establish their stability properties. Going over into 2D, we
present numerically-generated bifurcation diagrams for the fundamental and
vortex solitons, establish their stability ranges, and analyze a relation to
the 1D solution. To our knowledge, the existence of stable 2D solitons and
vortices supported by the cubic nonlinearity in the uniform space has not
been previously reported in any physical context. As concerns the
stabilization of 2D dissipative solitons by means of the feedback, provided
by a linearly coupled dissipative equation, this approach was first
proposed, in terms of anisotropic equations of the Kuramoto-Sivashinsky type
(its 2D modification), in Ref. \cite{Feng}. The model was suggested by
applications to the flow of viscous fluid films, rather than optics.

While our primary motivation is provided by the VCSEL-FSF, similar soliton
phenomena have been found in systems such as lasers with saturable gain and
absorption \cite%
{Vladimirov1997,Rosanov2005,RosanovPRL,Rosanov2003,Fedorov2003} and with a
holding beam \cite{LugiatoLefeverModel,FirthDamia2002,Damia2007,Damia2003},
as well as in VCSEL experiments employing coupled cavities \cite{Genevet2010}
and a built-in saturable absorber \cite{Elsass2010}, see also Ref. \cite%
{ackemann09} for a review.

The results reported in this work may have implications to both applied and
fundamental studies. On the one hand, the VCSEL-soliton systems offer a
strategy for the development of devices for all-optical information
processing applications. On the other hand, the results constitute a new
contribution to the great variety of dynamical phenomena described by CGL
systems in numerous physical contexts.

The article is organized as follows. In section \ref{sec_model} we present
the model previously considered for lasers with the FSF and show how it can
be reduced to a CGL3 equation coupled to a linear one. In section \ref%
{sec_zero_stability} we consider the stability of the zero solution, which
serves both as the background for self-localized modes and the source of
pattern-forming instabilities. We then review the dynamics obtained from
direct simulations of the CGL3 reduced model, which reproduces the
spontaneous 2D-soliton formation, which was reported, in terms of the full
model, in Ref. \cite{Paulau2008}, thus justifying the use of the reduction
to the cubic nonlinearity. In section \ref{sec_1Dsoliton} we discuss the
analytical 1D solution of our CGL3 model and produce a bifurcation diagram
similar to those found in more complex models \cite%
{Paulau2008,Paulau2009,Paulau2010}.

In section \ref{sec_2Dsoliton} we report the most essential new findings for
2D solitons. First, we extend the 1D analytic solution into that for the 2D
stripe-soliton family and describe this family, with an intention to
identify the stripe soliton as a limiting case of vortices. Then, using
polar coordinates in the 2D plane, we find and characterize a family of
vortex solitons, whose radius increases linearly with the topological
charge.
The stability of the vortex solitons is also analyzed in this section, both
for the restricted class of cylindrically-symmetric perturbations and full
azimuthal perturbations. While vortex solitons with high values of
topological charges $m$ are always subject to the azimuthal instability (in
particular, for $m\rightarrow \infty $ it goes over into the longitudinal
instability of the stripe), stable vortices with $|m|\leq 3$ are found in
our CGL3 model. The paper is concluded by section \ref{sec_summary}.


\section{The system and model}

\label{sec_model}

Following Ref. \cite{Paulau2010}, we start from the model for the
description of VCSELs coupled to frequency-selective feedback without delay:
\begin{equation}
\left\{
\begin{array}{l}
\frac{\partial E}{\partial t}=-\kappa (1+i\alpha )E+\kappa (1+i\alpha )\mu
\frac{E}{1+|E|^{2}}- \\
\qquad \qquad \qquad -i\Delta _{\perp }E+F-i\omega _{m}E, \\
\frac{\partial F}{\partial t}=-\lambda F+\sigma \lambda E,%
\end{array}%
\right.  \label{modeleqsSaturable}
\end{equation}%
where $\kappa $ is the cavity decay rate, $\alpha $ is the phase-amplitude
coupling factor, $\mu$ is the pump current, normalized to be $1$ at the
threshold in the absence of the external feedback, $\Delta _{\perp }$ is the
transverse Laplacian accounting for diffraction in the paraxial
approximation, $\omega _{m}$ is the detuning of the maximum of the
frequency-selective feedback profile from the laser's frequency at the
threshold without the feedback, $\lambda $ stands for the width of the
frequency filter, and $\sigma $ is the feedback strength in units of $\kappa
$, i.e., the threshold is reduced from $\mu =1$ at $\sigma =0$ to $\mu
=1-\sigma /\kappa $.

Truncating the Taylor expansion of the saturable nonlinearity at the third
order, Eq.~(\ref{modeleqsSaturable}) is approximated by a specific form of
the following CGL3 system
\begin{equation}
\left\{
\begin{array}{l}
\frac{\partial E}{\partial t}=g_{0}E+g_{2}|E|^{2}E+(d+iD)\Delta _{\perp }E+F,
\\
\\
\frac{\partial F}{\partial t}=-\lambda F+\tilde{\sigma}E,%
\end{array}%
\right.  \label{modeleqsGeneralized}
\end{equation}%
where
\begin{equation}
\begin{array}{l}
g_{0}=\kappa (1+i\alpha )(\mu -1)-i\omega _{m}, \\
g_{2}=-\kappa (1+i\alpha )\mu , \\
\tilde{\sigma}=\sigma \lambda, \\
D=-1, \\
d=0.%
\end{array}
\label{ParametersRenotation}
\end{equation}%
Note that $\mathrm{Re}(g_{0})$ is the total linear loss (if negative) or
gain (if positive), and $\mathrm{Im}(g_{0})$ plays the role of the effective
frequency detuning between the laser and the filter maximum. Further, $%
\mathrm{Re}(g_{2})$ is the nonlinear loss (if negative) or gain (if
positive), while $\mathrm{Im}(g_{2})$ represents the self-focusing or
defocusing nonlinearity. In the general case, real parameters $D$ and $d$
account for transverse diffraction and diffusion of the field. In the
present work we mainly consider $d=0$, which is relevant to optics models in
the spatial domain \cite{Mihalache2008,Herve1,Herve2,Herve3}, but it may be
different from zero in other physical situations -- in particular, in the
temporal domain \cite{Winful,Atai,Atai1998,Malomed2007}.

This approximation of the saturable nonlinearity by the cubic expansion is
justified by the fact that the higher-order nonlinearity, which usually
saturates the growth of the intensity in lasers without the feedback, is no
longer the main saturation mechanism in the presence of external
frequency-selective feedback. Above the threshold, the nonlinear term $i%
\mathrm{Im}(g_{2})|E|^{2}E$ induces a frequency shift that, together with
the frequency-dependent feedback, introduces an effective saturation capable
of limiting the field amplitude even without nonlinear losses [although $%
\mathrm{Re}(g_{2})$ is typically negative for lasers].

Model (\ref{modeleqsGeneralized}) is precisely the CGL3 equation coupled to
a linear equation. For laser models, there are usually specific relations
between $g_{0}$ and $g_{2}$; however, in other physical situations all
parameters of model (\ref{modeleqsGeneralized}) may be independent,
providing\ for the opportunity to study different behaviors of the solutions.

\section{Overview of the behavior of the system}

\label{sec_zero_stability}

Linearizing around the zero (non-lasing) solution, the evolution of
perturbations $\mathbf{\delta e}=\left( \delta E,\delta F\right) $ is
governed in the Fourier space by equation
\begin{equation}
\frac{d}{dt}\mathbf{\delta e}(k_{\perp })=\hat{M}(k_{\perp })\mathbf{\delta
e(k_{\perp })},  \label{stab}
\end{equation}
where
\begin{equation}
\hat{M}(k_{\perp })\equiv \left(
\begin{array}{lr}
g_{0}-(d+iD)k_{\perp }^{2} & 1 \\
\tilde{\sigma} & -\lambda%
\end{array}%
\right) .  \label{Mdef}
\end{equation}%
The stability of the zero solution against perturbations with wavenumber $k_{\perp }$ 
is determined by the eigenvalues of matrix $M$. Note that, since
the stability of the zero solution is independent of nonlinearities, the
analysis considered here is also valid for Eqs. (\ref{modeleqsSaturable})
for corresponding parameters.

\begin{figure}[tbp]
\begin{center}
\includegraphics[width=8cm,
keepaspectratio=true,clip=true]{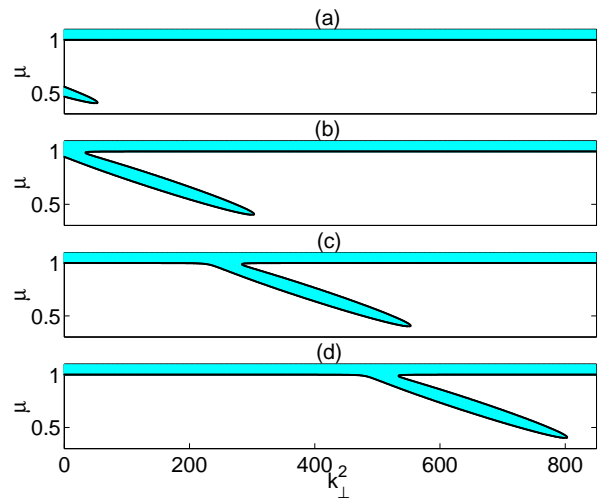}
\end{center}
\caption{(Color online). Shaded are regions of the instability of the zero
solution against perturbations with transverse wave number $k_{\perp }$ for
different values of detuning $\protect\omega _{m}$. The black line is the
marginal-stability boundary. (a) $\protect\omega _{m}=-250$, (b) $\protect%
\omega _{m}=0$, (c) $\protect\omega _{m}=250$, (d) $\protect\omega _{m}=500$%
. Other parameters are $\protect\sigma =60$, $\protect\kappa =100$, $\protect%
\lambda =2.71$, $\protect\alpha =5$, $d=0$.}
\label{fig_triv_stab_wm}
\end{figure}

\begin{figure}[tbp]
\begin{center}
\includegraphics[width=8cm,
keepaspectratio=true,clip=true]{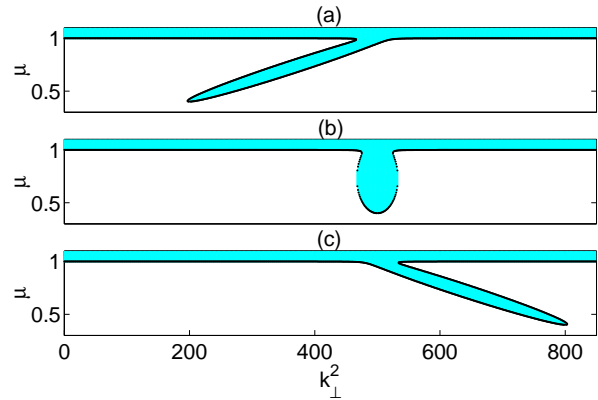}
\end{center}
\caption{(Color online). The same as in Fig.~\protect\ref{fig_triv_stab_wm}
for different values of $\protect\alpha $: (a) $\protect\alpha =-5$, (b) $%
\protect\alpha =0$, (c) $\protect\alpha =5$. Other parameters are $\protect%
\sigma =60$, $\protect\kappa =100$, $\protect\lambda =2.71$, $\protect\omega %
_{m}=500$, $d=0$.}
\label{fig_triv_stab_alpha}
\end{figure}

\begin{figure}[tbp]
\begin{center}
\includegraphics[width=8cm,
keepaspectratio=true,clip=true]{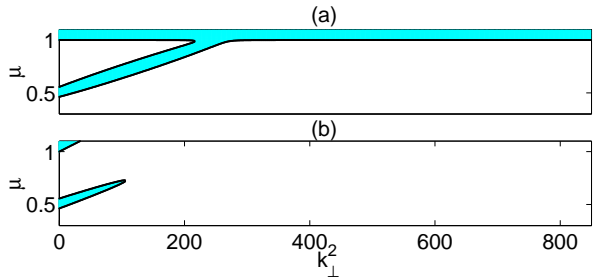}
\end{center}
\caption{(Color online). The same as in Fig.~\protect\ref{fig_triv_stab_wm}
for different values of $d$. Even in the self-defocusing case ($\protect%
\alpha <0$), the zero solution is stabilized by nonzero diffusion. Here $%
d=0.0$ (a) and $d=0.3$ (b). Other parameters are $\protect\sigma =60$, $%
\protect\kappa =100$, $\protect\lambda =2.71$, $\protect\alpha =-5$, $%
\protect\omega _{m}=250$.}
\label{fig_triv_stab_d}
\end{figure}

In Fig.~\ref{fig_triv_stab_wm} we show how the marginal stability curve
changes with detuning $\omega_{m}$. Taking into account the definition of $%
g_0$, for $d=0$ matrix $M$ depends on $\omega_m$ and $k_{\perp }$ only
through the combination $\omega_m+Dk^2_{\perp}$. Since $D=-1$, increasing $%
\omega_m$ the marginal stability curve translates rigidly to larger values
of $k^2_{\perp}$ as displayed in the figure. The mean slope of the
instability balloon depends on the value of $\alpha $ as shown in Fig.~\ref%
{fig_triv_stab_alpha}. An interesting property, exploited in previous works,
is the existence for positive $\alpha$ and for a range of negative values of
$\omega_m$ of a region of stability for the zero solution above the off-axis
emission threshold [See Fig.~\ref{fig_triv_stab_wm}(a)]. In this region,
stable self-localized modes can be found \cite{Paulau2008,Paulau2009}. In
contrast, for $\alpha <0$ [see Fig.~\ref{fig_triv_stab_alpha}(a)
corresponding to the self-defocusing case] there is no such region for any
value of $\omega _{m}$. The zero solution can be stabilized, though, by a
nonzero value of diffusion $d$ as illustrated by Fig.~\ref{fig_triv_stab_d}.
Note also that both for the self-focusing and defocusing nonlinearity the
zero solution can be stabilized by filtering of spatial Fourier modes in the
feedback loop, which can be modeled by a wavenumber-dependent feedback
strength $\tilde{\sigma}(k_{\perp})$\cite{Paulau2007}.

\begin{figure}[tbp]
\begin{center}
\includegraphics[width=8cm,
keepaspectratio=true,clip=true]{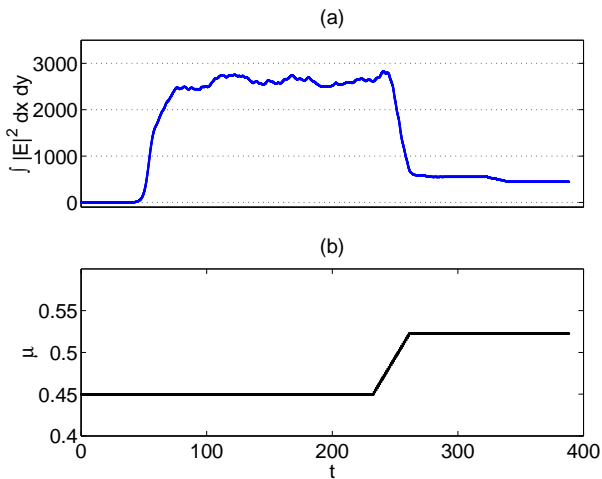}
\end{center}
\caption{(Color online). (a) The total intensity of the 2D laser field,
following the switch-on at the pump current $\protect\mu =0.45$, and then
the transition to a set of quiescent localized spots, achieved by ramping
the pump up to $\protect\mu =0.52$. (b) The variation of the pump current in
time which gives rise to the dynamical picture displayed in panel (a). The
parameters are: $\protect\sigma =60$, $\protect\kappa =100$, $\protect%
\lambda =2.71$, $\protect\alpha =5$, $\protect\omega _{m}=-270$, $d=0$.}
\label{fig_dyn}
\end{figure}

\begin{figure}[tbp]
\begin{center}
\includegraphics[width=8cm,
keepaspectratio=true,clip=true]{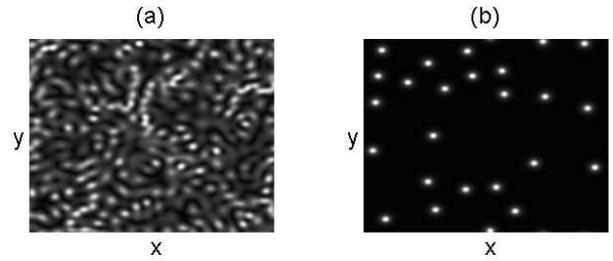}
\end{center}
\caption{The instantaneous 2D amplitude profile
corresponding to the regimes presented in Fig.~\protect\ref{fig_dyn}, at:
(a) $t=200$, corresponding to the pump current at which the zero solution is
unstable and a complex spatiotemporal regime arises; (b) $t=400$, after
having increased the pump current to a value for which the zero solution is
stable, and the spatiotemporal pattern decays into a set of fundamental
solitons.}
\label{fig_space}
\end{figure}

Different spatiotemporal regimes are possible in model (\ref%
{modeleqsGeneralized}) depending on values of the parameters. For a set of
parameters close to those corresponding to Fig.~\ref{fig_triv_stab_wm}(a),
which are relevant to the dynamics of lasers, the following scenario is
observed. For the pump currents in the unstable region ($\mu \approx 0.45$),
small random perturbations of the zero solution grow exponentially, see Fig.~%
\ref{fig_dyn}(a), leading to a complex 2D spatiotemporal pattern, which sets
in at $t\approx 80$ in Fig.~\ref{fig_dyn}(a). The instantaneous spatial
profile of this chaotic pattern is shown in Fig.~\ref{fig_space}(a). If,
starting from this regime, pump current $\mu $ is further increased up to a
value at which the zero background is stable ($\mu =0.52$) [see Fig.~\ref%
{fig_dyn}(b)], a transition is observed from densely packed filaments to a
2D set of isolated quiescent solitons, see Fig.~\ref{fig_space}(b).
Initially, the distance between the solitons is small, and the interaction
among them leads to a partial annihilation, see the abrupt fall of the
integral intensity at $t\approx 350$ in Fig.~\ref{fig_dyn}(a). However, when
the density of solitons becomes small enough, the resulting set of quiescent
cavity solitons is quasi-stationary, featuring very weak interactions.

Here we have presented the results for $\omega _{m}=-270$, for which the
marginal stability curve is the same as in Fig.~\ref{fig_triv_stab_wm}(a)
but displaced 20 units to the left. For values of $\omega_{m}\approx-250$
moving solitons instead of quiescent ones are observed in the final state.

A similar scenario is observed in one dimension, 
which is of special interest because exact 1D self-localized solutions are
available in the CGL3 system, as we discuss below.

\section{One-dimensional solitons}

\label{sec_1Dsoliton}

In the case of one transverse dimension ($\Delta _{\perp }=\partial
^{2}/\partial x^{2}$) an exact analytical solution to Eqs. (\ref%
{modeleqsGeneralized}) can be found in the form of \cite%
{Atai1998,Malomed2007,Firth2010}
\begin{equation}
\left\{
\begin{array}{l}
E=E_{\max }\left[ \cosh (Kx)\right] ^{-1-i\beta }e^{i\omega t}, \\
F=F_{\max }\left[ \cosh (Kx)\right] ^{-1-i\beta }e^{i\omega t}.%
\end{array}%
\right.  \label{analyt_sol}
\end{equation}%
Substituting expressions (\ref{analyt_sol}) into Eqs. (\ref%
{modeleqsGeneralized}), we eliminate
\begin{equation}
F_{\max }=\frac{\tilde{\sigma}}{\lambda +i\omega} E_{\mathrm{\max}},
\label{fmax}
\end{equation}%
and obtain the following quadratic equation for chirp $\beta $,
\begin{equation}
\beta ^{2}+3\beta \frac{\mathrm{Re}\left[ g_{2} (d-iD)\right] }{\mathrm{Im}%
\left[g_{2}(d-iD)\right] }-2=0,  \label{ChirpEquation}
\end{equation}%
which yields a single physical root, due to the condition that the field
intensity
\begin{equation}
|E_{\mathrm{\max }}|^{2}=3\beta K^{2} \frac {d^2+D^2} {\mathrm{Im}\left[%
g_{2}(d-iD)\right]}  \label{IntensityEquation}
\end{equation}%
must be positive. Note that $\beta $ does not depend on linear coefficient $%
g_{0}$, but only on nonlinear coefficient $g_{2}$ and on the parameters of
spatial coupling, $d$ and $D$. Once $\beta $ is known, a complex algebraic
equation involving $\omega $ and $K$ is obtained. By separating the real and
imaginary parts of this equation
we obtain
\begin{equation}
K^{2}=-\frac{\mathrm{Re}(g_{0})}{\mathrm{Re}\left( \tilde{\beta}\right) }-%
\frac{\tilde{\sigma}\lambda }{\lambda ^{2}+\omega ^{2}}\frac{1}{\mathrm{Re}%
\left( \tilde{\beta}\right) },  \label{KEquation}
\end{equation}
where
\begin{equation}
\tilde{\beta}\equiv (1+i\beta )^{2}(d+iD).  \label{auxiliarybeta}
\end{equation}
Next, we obtain a cubic equation for $\omega $:
\begin{equation}
a_{3}\omega ^{3}+a_{2}\omega ^{2}+a_{1}\omega +a_{0}=0,
\label{FrequencyEquation}
\end{equation}%
where coefficients $a_{0}$, $a_{1}$, $a_{2}$, and $a_{3}$ depend on the
system's parameters and on $\tilde{\beta}$:
\begin{eqnarray}
\begin{array}{l}
a_{3}=\mathrm{Re}(\tilde{\beta}), \\
a_{2}=\mathrm{Re}(g_{0})\mathrm{Im}(\tilde{\beta})-\mathrm{Im}(g_{0})\mathrm{%
Re}(\tilde{\beta}), \\
a_{1}=(\tilde{\sigma}+\lambda ^{2})\mathrm{Re}(\tilde{\beta}), \\
a_{0}=a_{2} \lambda ^{2}+\tilde{\sigma}\lambda \mathrm{Im}(\tilde{\beta}).%
\end{array}
\label{FrequencyCoefficients}
\end{eqnarray}
Equation (\ref{FrequencyEquation}) can be solved analytically, but it is
more practically relevant to solve it numerically, as in Ref. \cite%
{Firth2010}. To summarize, the analytical solutions can be constructed
according to the following scheme.
\begin{list}{$\bullet$}{}
\item (i) Solve Eq. (\ref{ChirpEquation}) for $\beta$, and choose the proper
root to satisfy the positivity of $|E_{max}|^2$ as per Eq.
(\ref{IntensityEquation}).
\item (ii) Solve Eq. (\ref{FrequencyEquation}) for $\omega$ with the coefficients
defined by Eqs. (\ref{auxiliarybeta}, \ref{FrequencyCoefficients}),
and $\beta$ produced by the previous step.
\item (iii) Calculate $K$ using Eqs. (\ref{KEquation}, \ref{auxiliarybeta}), and $\omega$,
with $\beta$ produced by two previous steps.
\item (iv) Calculate $E_{max}$ using Eq. (\ref{IntensityEquation}).
\item (v) Calculate  $F_{max}$ using Eq. (\ref{fmax}).
\end{list}
Once all parameters of solution (\ref{analyt_sol}) are determined, the
stability of this solution can be analyzed following the numerical procedure
developed in Ref. \cite{Paulau2010}.

\begin{figure}[tbp]
\begin{center}
\includegraphics[width=8cm,
keepaspectratio=true,clip=true]{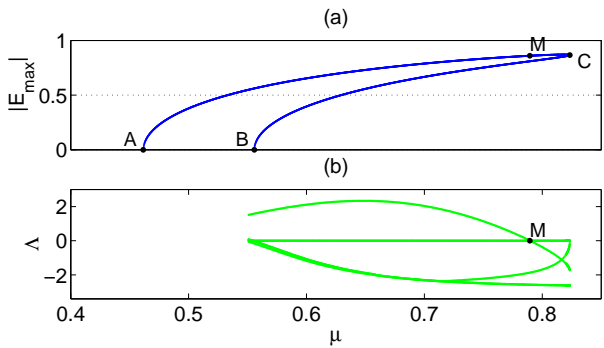}
\end{center}
\caption{(Color online). The bifurcation diagram for analytical solution (%
\protect\ref{analyt_sol}). (a) The amplitude (absolute value of the field at
the center of the fundamental soliton) as a function of the pump current.
(b) The growth rate of unstable, neutral and least-damped stable
perturbation eigenmodes versus the pump current $\protect\mu $. Point $M$
designates the drift instability of the soliton. The parameters are $\protect%
\omega _{m}=-250$, $\protect\alpha =5$, $\protect\kappa =100$, $\protect%
\sigma =60$, $\protect\lambda =2.71$.}
\label{fig_horn_stab}
\end{figure}

Fig.~\ref{fig_horn_stab}(a) shows the bifurcation diagram for the localized
quiescent solitons (\ref{analyt_sol}), using parameter values typical for
the lasers models. Point A corresponds to the pump threshold for on-axis
emission. For the pump levels $\mu$ between points A and B the zero
background is unstable as shown in Fig.~\ref{fig_triv_stab_wm}(a). These
points coincide with the origin of two branches of localized structures. The
two branches collide at C and disappear through a saddle-node bifurcation.

Figure~\ref{fig_horn_stab}(b) shows the real part of the most relevant
eigenvalues resulting of the stability analysis of the upper branch [the one
connecting points C, M and A in Fig.~\ref{fig_horn_stab}(a)]. The soliton
solution is stable between points C and M, and a drift instability appears
at point M \cite{Paulau2009}. The instability spectrum is not shown between
points A and B because the background zero solution is unstable. The lower
branch of soliton solution connecting points B and C in Fig.~\ref%
{fig_horn_stab}(a) is entirely unstable, as usual \cite{Winful,Atai}.

\begin{figure}[tbp]
\begin{center}
\includegraphics[width=8cm,
keepaspectratio=true,clip=true]{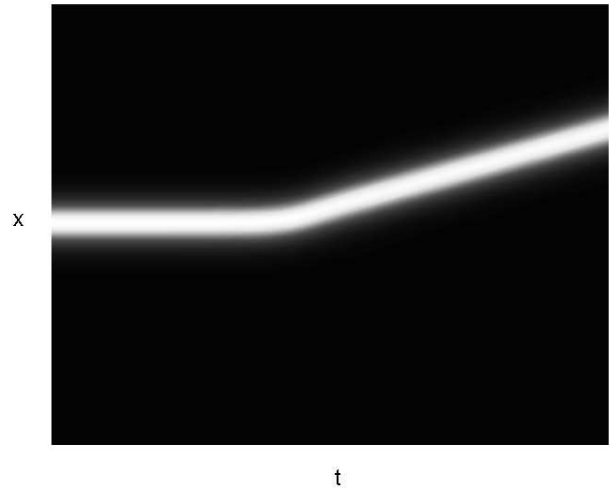}
\end{center}
\caption{Spatiotemporal dynamics of the 1D soliton in the
course of the development of the drifting instability. The parameters are as
in Fig.~\protect\ref{fig_horn_stab}, for $\protect\mu =0.7$.}
\label{fig_1Ddrift}
\end{figure}

Direct simulations starting from the 1D analytic soliton in the unstable
region (to the left of point M in Fig.~\ref{fig_horn_stab}) shows the
development of the drift instability, see Fig.~\ref{fig_1Ddrift}. Notice
that once the drift instability sets in the soliton moves away from its
original location at a constant speed.

The overall scenario is very similar to that found in previous numerical
works for the saturable nonlinearity \cite{Paulau2008,Paulau2009,Paulau2010}%
, suggesting that the present CGL3 system indeed represents a simple
underlying model which captures the essential features of more realistic,
but also more involved, models.

\section{Two-dimensional self-localized solutions: stripes, fundamental, and
vortex solitons}

\label{sec_2Dsoliton}

\begin{figure}[tbp]
\begin{center}
\includegraphics[width=8cm,
keepaspectratio=true,clip=true]{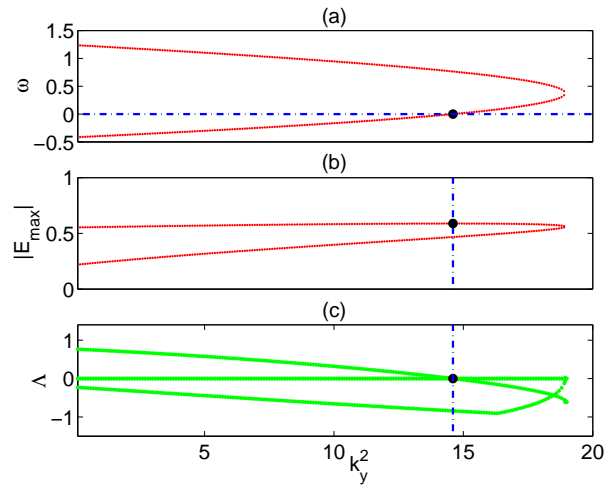}
\end{center}
\caption{(Color online). The continuous family of 2D stripe solitons
parameterized by $k_{y}$, for $\protect\mu =0.52$ and other parameters taken
as in Fig.~\protect\ref{fig_dyn}. (a) Frequency $\protect\omega $, (b) the
soliton's amplitude $|E_{\max }|$, and (c) growth rates of the unstable
neutral and least damped stable eigenmodes (the stability is considered
against perturbations depending only on the $x$ coordinate).}
\label{fig_2D_stripe_soliton_family}
\end{figure}

In 2D ($\Delta _{\perp }=\frac{\partial ^{2}}{\partial x^{2}}+\frac{\partial
^{2}}{\partial y^{2}}$), the 1D solution (\ref{analyt_sol}) can be
generalized to a continuous family of the stripe-soliton solutions,
parameterized by transverse wavenumber $k_{y}$:
\begin{equation}
\left\{
\begin{array}{l}
E=E_{\max }\left[ \cosh (Kx)\right] ^{-1-i\beta }e^{i\omega t}e^{ik_{y}y},
\\
F=F_{\max }\left[ \cosh (Kx)\right] ^{-1-i\beta }e^{i\omega t}e^{ik_{y}y}.%
\end{array}%
\right.  \label{analyt_stripesol}
\end{equation}%
The only difference from the above 1D solution is a modification of linear
coefficient $g_{0}$, which is replaced by $\tilde{g}%
_{0}=g_{0}-(d+iD)k_{y}^{2}$. This solution is shown in Fig.~\ref%
{fig_2D_stripe_vortex_soliton}(a,b). We have found the whole family of the
stripe solitons as a function of $k_{y}$, see Fig.~\ref%
{fig_2D_stripe_soliton_family}. They exist only for values of $k_y^2$ below
a certain value beyond which the frequency shift introduced by $k_{y}$
pushes the solution outside the frequency range of the feedback filter. The
solution with largest amplitude, marked by a filled circle, corresponds to $%
k_{y}$ such that $\omega=0$. Fig.~\ref{fig_2D_stripe_soliton_family}(c),
shows the largest real parts of the eigenvalues obtained from the linear
stability analysis in the $x$-direction. The stripe-soliton undergoes a
drift instability similar to the one of the 1D soliton described in the
previous section. Here the stripe as a whole would start to move either to
the left or to the right of its axis. Interestingly enough the drift
instability takes place at a value of $k_y=k_y^c$ which, within the
numerical accuracy, coincides with the value for which the solution has $%
\omega=0$. The critical value $k_y^c$ will be very relevant later when
studying the radial dynamics of vortices. In any case, 2D stripe-solitons
are always unstable to perturbations in the $y$-direction, breaking up into
a number of fundamental (spot) solitons.

\begin{figure}[tbp]
\begin{center}
\includegraphics[width=8cm,
keepaspectratio=true,clip=true]{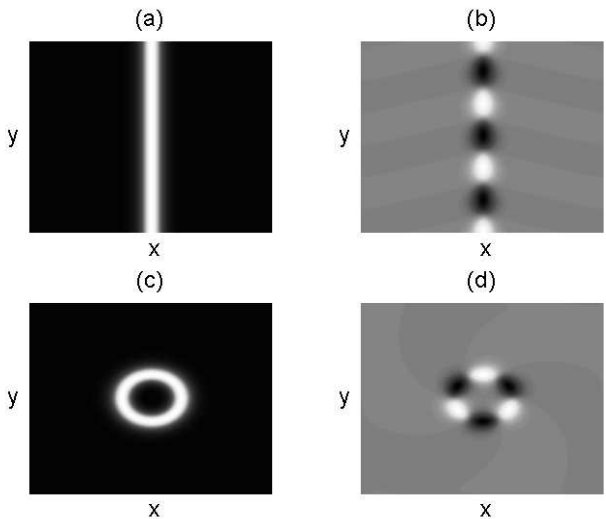}
\end{center}
\caption{The spatial profile of the amplitude (a) and real
part of the field (b) for the unstable 2D stripe soliton (\protect\ref%
{analyt_stripesol}). (c) and (d): The same as (a) and (b) for a stable
vortex with $m=3$. Here $\protect\mu =0.52$ and other parameters are the
same as in Fig.~\protect\ref{fig_dyn}.}
\label{fig_2D_stripe_vortex_soliton}
\end{figure}

We now proceed to fully localized 2D solutions. There are two types of
stable 2D modes: fundamental solitons with the bell-like intensity profile,
see Fig. \ref{fig_space}(b),
%
%
and ring-shaped vortex solitons, see Fig.~\ref{fig_2D_stripe_vortex_soliton}%
(c). Vortex solitons with integer topological charge $m$ can be looked for
as $E(r,\phi )=E_{0}(r)e^{im\phi }$, where $(r,\phi )$ are polar coordinates
with the origin at the pivot of the vortex \cite%
{Montagne1997,CrasovanMalomedMihalache2000,Mihalache2008,Desyatnikov2000,Towers2001,Herve3}%
. The fundamental 2D soliton corresponds to $m=0$, with the maximum at the
origin. Every $m$ vortex has a mirror-image $-m$ vortex, therefore for the
sake of simplicity in what follows we will consider vortex solitons with $%
m>0 $.

Figure~\ref{fig_2D_stripe_vortex_soliton} shows the similarity between a
vortex mode and the stripe soliton. Roughly speaking, the vortex may be
considered as a stripe bent into a closed circle, at least for large values
of $m$. Following this similarity, we study the radial dynamics of vortices
using the radial version of Eq.~(\ref{modeleqsGeneralized}), with
\begin{equation}
\Delta _{\perp }=\frac{\partial ^{2}}{\partial r^{2}}+\frac{1}{r}\frac{%
\partial }{\partial r}-\frac{m^{2}}{r^{2}}.  \label{RadialLaplacian}
\end{equation}

\begin{figure}[h]
\begin{center}
\includegraphics[width=8cm,
keepaspectratio=true,clip=true]{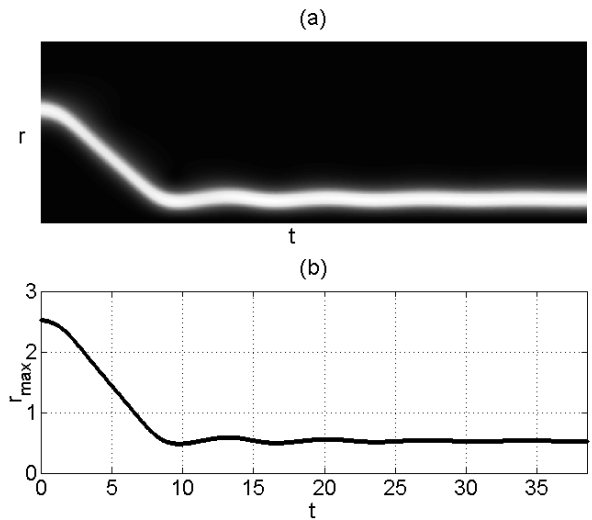}
\end{center}
\caption{(a) The spatiotemporal dynamics of the radial
profile of the field amplitude relaxing towards the equilibrium radius of
the vortex with $m=2$. Here $\protect\mu =0.52$ and other parameters are as
in Fig.~\protect\ref{fig_dyn}. (b) The dynamics of position $r_{\max }$ of
the maximum of the field. }
\label{fig_rdyn}
\end{figure}

Using 1D analytical solution (\ref{analyt_sol}) to define the initial
condition as $E_{\mathrm{ini}}(r)=E_{0}(r=R_{0}+x)$, we simulated Eq. (\ref%
{modeleqsGeneralized}) with the Laplacian taken as per Eq. (\ref%
{RadialLaplacian}), with fixed $m$. If the initial ring radius $R_{0}$ is
too small, the field decays to zero%
, but for $R_{0}$ large enough we observed the evolution of radius $r_{\max}
$ corresponding to the maximum field amplitude towards equilibrium radius $%
R^{\mathrm{st}} (m)$ of the vortex soliton of charge $m$, see Fig.~\ref%
{fig_rdyn}. If $R_{0}$ is much larger than $R^{\mathrm{st}} (m)$, we
observed that the vortex shrank at a constant speed, which was followed by
relaxation oscillations as $R^{\mathrm{st}} (m)$ was approached. Actually,
this is an unusual behavior, very different from the typical
curvature-driven dynamics \cite{Denmark,Gomila01}. The constant shrinkage
speed may be explained by considering the quasi-1D dynamics of the
stripe-soliton solution.
If the initial condition has a very large $r_{\max }$, the ring can be
considered locally as a stripe-soliton with $k_{y} \approx 0$, which is
drift-unstable (see Fig.~\ref{fig_2D_stripe_soliton_family}). The overall
curvature of the ring breaks the left-right symmetry the stripe had in the
direction perpendicular to the axis. The symmetry breaking is such that the
drift takes place towards the center and hence the ring as a whole starts
contracting at a constant speed. The first stage of the dynamics shown in
Fig.~\ref{fig_rdyn} (at $t<7$) is, thus, essentially the same as in Fig.~\ref%
{fig_1Ddrift}. As radius $r_{\max }$ shrinks, the effective wavenumber $%
k_{y} $ increases, eventually suppressing the drift instability and the ring
relaxes to the stable radius $R^{\mathrm{st}} (m)$.
%
%

\begin{figure}[tbp]
\begin{center}
\includegraphics[width=8cm,
keepaspectratio=true,clip=true]{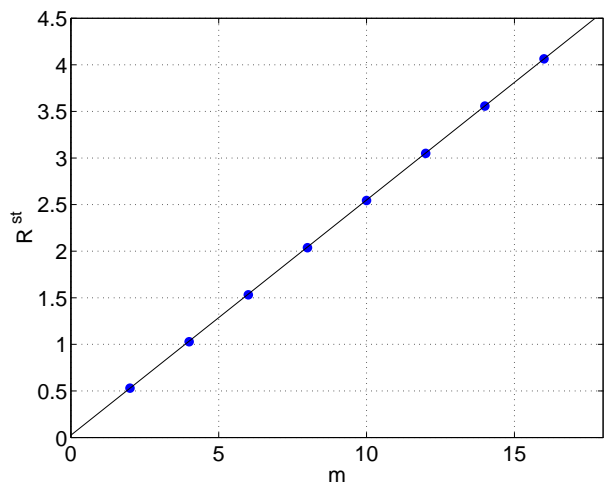}
\end{center}
\caption{(Color online). The equilibrium radius $R^{\mathrm{st}}$ of the vortex solitons
versus the topological charge. Parameters are as in Fig.~\protect\ref%
{fig_rdyn}. Filled circles mark even integer values of $m$, while the solid
line is a linear fitting. Figure~\protect\ref{fig_rdyn} corresponds to the
leftmost point in this figure.}
\label{fig_radiusm}
\end{figure}

Running the simulations for different (large enough) values of $m$, we have
produced the dependence of the equilibrium radius $R^{\mathrm{st}}$ on the
topological charge $m$, which turns out to be linear, see Fig. \ref%
{fig_radiusm}. Therefore, vortex rings expand as $m$ increases, tending
towards the stripe solution in the limit of $m\rightarrow \infty $. The
inverse of the slope of the line in Fig. \ref{fig_radiusm}, $R^{\mathrm{st}}
(m)/m$, is the transverse circular wavenumber $k_{c}$, which is, evidently,
nearly constant for all vortices. The value of $k_{c}$ turns out be very
similar to the drift-instability critical wavenumber of the soliton stripe, $%
k_{y}^c$ considered above.

The mechanism leading to 2D stable vortices discussed here has no
counterpart in simple curvature driven dynamics of fronts connecting two
equivalent states \cite{Denmark,Gomila01}. In these systems, 1D fronts in a
2D system may be subject to modulational instabilities but not to drift
ones. Therefore there is no transient regime in which the ring radius
changes at a constant rate. The existence of the 1D soliton drift
instability plays a critical role in the dynamics of 2D solitons and
determines its stationary size.

The radial equation allows one to study the radial dynamics independently of
the presence of azimuthal instabilities. In fact, as in the case of the
stripe soliton, vortices with large $m$ are azimuthally unstable in the full
2D problem. The curvature can, however, prevent the azimuthal instability
for small topological charges, and vortices may be stable up to a certain
value of $m$ \cite{CrasovanMalomedMihalache2000,Paulau2010}. Figure~\ref%
{fig_2D_stripe_vortex_soliton}(c,d) shows, for instance, a stable vortex for $m=3$.

\begin{figure}[tbp]
\begin{center}
\includegraphics[width=8cm,
keepaspectratio=true,clip=true]{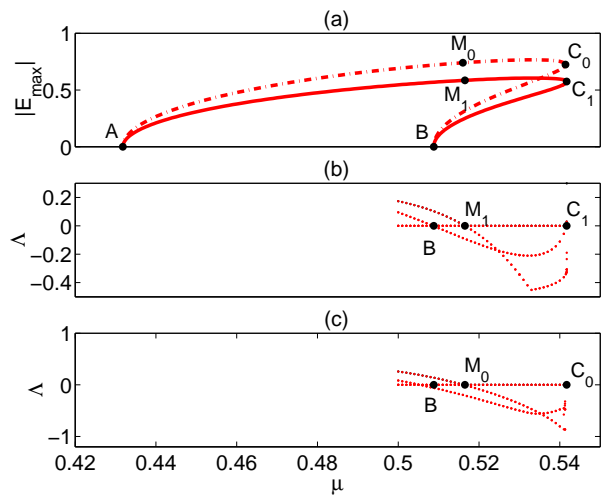}
\end{center}
\caption{(Color online). (a) The amplitude of the soliton with $m=0$
(dash-dotted line) and $m=1$ (solid line) versus the pump current. (b) The
growth rate of the 6 localized modes with highest $\Lambda$, including
unstable, neutral and least damped stable eigenmodes for the solution branch
C$_{1}$M$_{1}$A in panel (a). (c) The same as (b) but for the solution
branch C$_0$M$_{0}$A in panel (a). Parameters are the same as used above in
this section.}
\label{fig_m1m0br2D270}
\end{figure}

Finally, following the method described in Ref. \cite{Paulau2010}, we have
computed the bifurcation diagrams of the solitons with $m=0,1$ and analyzed
their stability, see Fig. \ref{fig_m1m0br2D270}. The fundamental soliton
(vortex) is stable between points $M_{0}$ and $C_{0}$ ($M_{1}$ and $C_{1}$).
Point $M_{0} (M_{1})$ again corresponds to the onset of the drifting
instability of the state as a whole. The branches connecting points $B$ and $%
C_0$ ($B$ and $C_1$) correspond to the solitons which play the role of the
unstable separatrix. The fundamental 2D soliton and the $m=1$ vortex have
quite a similar bifurcation diagram. As compared to 1D solitons, see Fig. %
\ref{fig_horn_stab}, points A and B correspond to the limits of the region
where the background zero solution is unstable to homogeneous perturbations
and therefore are the same, however here points C and M are located at a
lower pump value.

In addition, we have observed stable vortices with $m=2$ and $3$. The region
of their existence is almost identical to the existence region of $m=1$
vortex, while the stability region is narrower and lies inside M$_1$C$_1$
interval of Fig.~\ref{fig_m1m0br2D270}.

In the case of saturable nonlinearity, system (\ref{modeleqsSaturable})
gives also rise to fundamental 2D solitons and vortices as encountered here
\cite{Paulau2010}. The region of existence and the bifurcation diagrams are
quite similar to the ones shown here. This is a clear indication that the
present CGL3 model is indeed relevant for other systems, for which the
analysis in terms of exact stripe solutions is not possible.

\section{Summary}

\label{sec_summary}

In this work, we have introduced the system of coupled cubic complex
Ginzburg-Landau equation and additional dissipative linear equation as the
model of laser cavities with the external frequency-selective feedback.
We have observed a qualitative agreement with the results recently obtained
in models with the saturable nonlinearity \cite%
{Paulau2008,Paulau2009,Paulau2010}. In particular, the stability of the
fundamental 2D solitons is obvious in the case of the saturable
nonlinearity, while it is a nontrivial finding in the cubic model. Using
analytical considerations and numerical analysis, we have shown that 2D
vortex solitons can be interpreted as stripe solitons bent into rings (as
illustrated by Fig. \ref{fig_2D_stripe_vortex_soliton}). This correspondence
is clear for $m\rightarrow \infty $, but it actually holds too for rather
small $m$, since the circular transverse wavenumber appears to be the same
for all $m$ (see Fig.~\ref{fig_radiusm}). In such a way, we have established
the connection between 1D and 2D solitons.

In our system of the coupled cubic and linear equations, we have found
stable 1D and 2D solitons, including vortices. These results are important,
as they show that this system may be considered as the fundamental model
underlying a wide class of stable soliton lasers. The simplicity and
flexibility of the linear coupling has previously been shown to provide the
existence of stable 1D solitons in the models of dual-core waveguides \cite%
{Winful,Atai,Atai1998,Sakaguchi,Malomed2007}. Our extension of this approach
into the spatial domain, and into 2D, means that models of this class may be
useful to describe and analyze stable self-localization for a wide variety
of physical systems.

We acknowledge financial support from MICINN (Spain) and FEDER (EU) through
Grant No. FIS2007-60327 FISICOS.

\vspace{\parindent}

\clearpage
\newpage

\end{document}